\documentclass[journal, twoside]{IEEEtran}
\usepackage{indentfirst}
\usepackage{graphicx}
\usepackage{amsmath}
\usepackage{amssymb}
\usepackage{amsfonts}
\usepackage{mathrsfs}
\usepackage{leftidx}
\usepackage{color}
\usepackage{amsmath}
\usepackage{arydshln}
\usepackage{amsthm}
\usepackage{ragged2e}
\usepackage{cite}
\usepackage{enumerate}
\usepackage{longtable}
\usepackage{float}
\usepackage{stfloats}
\usepackage{hyperref}
\usepackage{algpseudocode}
\usepackage{algorithm}
\usepackage[caption=false,font=footnotesize]{subfig}
\usepackage{tabularx}
\usepackage{makecell}
\usepackage{url}
\usepackage{tabularx}

\usepackage{multirow} 
\usepackage{booktabs}
\theoremstyle{plain}

\usepackage{caption}
\usepackage[table]{xcolor}
\usepackage{multirow}
\usepackage{array}

\newcolumntype{P}[1]{>{\raggedright\arraybackslash\footnotesize}m{#1}}
\newcolumntype{A}[1]{>{\centering\arraybackslash\footnotesize}m{#1}}

\usepackage[table,usenames,dvipsnames]{xcolor}
\definecolor{aa}{RGB}{175,238,238}
\definecolor{bb}{RGB}{255,255,255}

\usepackage{bm}
\usepackage{makecell}

\begin{document}

\title{Intellicise Wireless Networks Meet Agentic AI: A Security and Privacy Perspective}

\author{Rui Meng,~\IEEEmembership{Member,~IEEE,} Zhidi Zhang, Song Gao, Yaheng Wang, Xiaodong Xu,~\IEEEmembership{Senior Member,~IEEE,} 

Yijing Lin,~\IEEEmembership{Member,~IEEE,} 
Yiming Liu,~\IEEEmembership{Member,~IEEE,} Chenyuan Feng,~\IEEEmembership{Member,~IEEE,} 

Lexi Xu,~\IEEEmembership{Senior Member,~IEEE,} 
Yi Ma,~\IEEEmembership{Senior Member,~IEEE,} 
Ping Zhang,~\IEEEmembership{Fellow,~IEEE,} 

and Rahim Tafazolli,~\IEEEmembership{Fellow,~IEEE} 

\thanks{
\textit{(Corresponding author: Rui Meng and Xiaodong Xu.)}

Rui Meng, Zhidi Zhang, Song Gao, Yaheng Wang, Xiaodong Xu, Yijing Lin, Yiming Liu, and Ping Zhang are the State Key Laboratory of Networking and Switching Technology, Beijing University of Posts and Telecommunications, Beijing, China (e-mail: buptmengrui@bupt.edu.cn; 2639134068@bupt.edu.cn; wkd251292@bupt.edu.cn; wangyaheng@bupt.edu.cn; xuxiaodong@bupt.edu.cn; yjlin@bupt.edu.cn; liuyiming@bupt.edu.cn; pzhang@bupt.edu.cn).

Chenyuan Feng is with Department of Computer Science, University of Exeter, EX4 4QF Exeter, U.K. (e-mail: c.feng@exeter.ac.uk).

Lexi Xu is with the Research Institute, China United Network Communications Corporation, Beijing, China (e-mail: davidlexi@hotmail.com).

Yi Ma and Rahim Tafazolli are with 5GIC \& 6GIC, Institute for Communication Systems (ICS), University of Surrey, Guildford, GU2 7XH, United Kingdom (email: y.ma@surrey.ac.uk; r.tafazolli@surrey.ac.uk).

}}

\maketitle

\begin{abstract}
\textit{Intellicise (Intelligent and Concise)} wireless network is the main direction of the evolution of future mobile communication systems, a perspective now widely acknowledged across academia and industry. As a key technology within it, Agentic AI has garnered growing attention due to its advanced cognitive capabilities, enabled through continuous perception-memory-reasoning-action cycles. This paper first analyses the unique advantages that Agentic AI introduces to intellicise wireless networks. We then propose a structured taxonomy for Agentic AI-enhanced secure intellicise wireless networks. Building on this framework, we identify emerging security and privacy challenges introduced by Agentic AI and summarize targeted strategies to address these vulnerabilities. A case study further demonstrates Agentic AI's efficacy in defending against intelligent eavesdropping attacks. Finally, we outline key open research directions to guide future exploration in this field.
\end{abstract}


\section{Introduction}

To address the growing diversity of service demands in mobile communication systems while optimizing the use of limited network resources, \textit{Intellicise (intelligent and concise)} wireless networks integrate foundational theories such as information theory, artificial intelligence (AI) theory, and systems theory, thus achieving intention-driven, semantic bearing, and distributed autonomy capabilities \cite{zhang2025intellicise,fan2025generative}. The integration also introduces new security issues. For instance, sensing information such as Channel State Information (CSI) is used to enhance communication and network performance, but intelligent attackers may capture CSI samples from public environments and analyze private information about users' daily behaviors. Additionally, intelligent eavesdroppers can utilize advanced signal processing and decoding techniques to reconstruct private semantic information transmitted in public environments \cite{meng2025image}. 

In response to security and privacy threats in intellicise wireless networks, generative AI leverages its strengths in unsupervised learning and diverse content generation to address challenges faced by discriminative AI-based defense technologies. However, static generative AI-based defenses face the following challenges. First and foremost, they exhibit inability to actively defend: relying on manual input to trigger feedback mechanisms, they cannot proactively set defense goals or define defense processes. Furthermore, they encounter difficulty in coping with dynamic attacks: while capable of generating security analysis reports, they fail to directly translate these into defensive actions, necessitating human intervention for final responses. Ultimately, they reveal a lack of continuous learning and evolutionary capabilities: struggling to dynamically optimize models based on real-time attack-defense feedback, they depend on manual data re-injection and model updates to counter emerging threats.

Against these challenges, Agentic AI, intelligent systems characterized by continuous perception-memory-reasoning-action loops, facilitates autonomous attack and threat perception, defense reasoning, and defense decision-making \cite{zhang2026toward}. 
Firstly, Agentic AI achieves proactive defense through a perception-driven autonomous decision-making loop. Secondly, leveraging techniques such as multi-agent deep reinforcement learning (DRL), game theory, and meta-learning, Agentic AI constructs dynamic response models to enable real-time collaborative defense. Thirdly, by employing techniques like retrieval-augmented generation (RAG) to update attack samples and defense strategies, Agentic AI achieves sustained learning and evolutionary defense enhancement \cite{jiang2025large}.
Nevertheless, as an emerging technique, Agentic AI may introduce additional threats to intellicise wireless networks \cite{datta2025agentic}.
For example, multi-agent collusion creates erroneous group consensus, coupled with premature truncation of logical chains, thus undermining the self-evolutionary and objective-driven reasoning within Intellicise wireless networks \cite{raza2025trism}.
Therefore, how to defend against the potential security risks is also important.

Against this background, we present a systematic investigation into integrating intellicise wireless networks and Agentic AI from a security and privacy perspective. The primary contributions are outlined as:
\begin{itemize}
\item We analyze the benefits that Agentic AI brings to intellicise wireless networks, emphasizing its transformative potential in enhancing semantic representation, network optimization, and scenario performance.
\item We propose a structured taxonomy for Agentic AI-enhanced secure intellicise wireless networks, encompassing three key domains: secure intellicise signal processing, secure intellicise information transmission, and secure intellicise network organization.
\item We identify the security and privacy issues introduced by Agentic AI in intellicise wireless networks, and propose targeted strategies to secure Agentic AI-enhanced implementations, addressing emerging threats and vulnerabilities.
\item We present a case study demonstrating that Agentic AI can effectively defend against intelligent eavesdropping attacks in intellicise wireless networks.
\item We outline several open issues to guide subsequent investigations in this field.
\end{itemize}

\section{Overview of Agentic AI-enhanced Intellicise Wireless Networks}

In this section, we first introduce Intellicise wireless networks and Agentic AI, followed by an exploration of several key junctions between them.

\begin{figure*}
\centering
\includegraphics[width=1\textwidth]{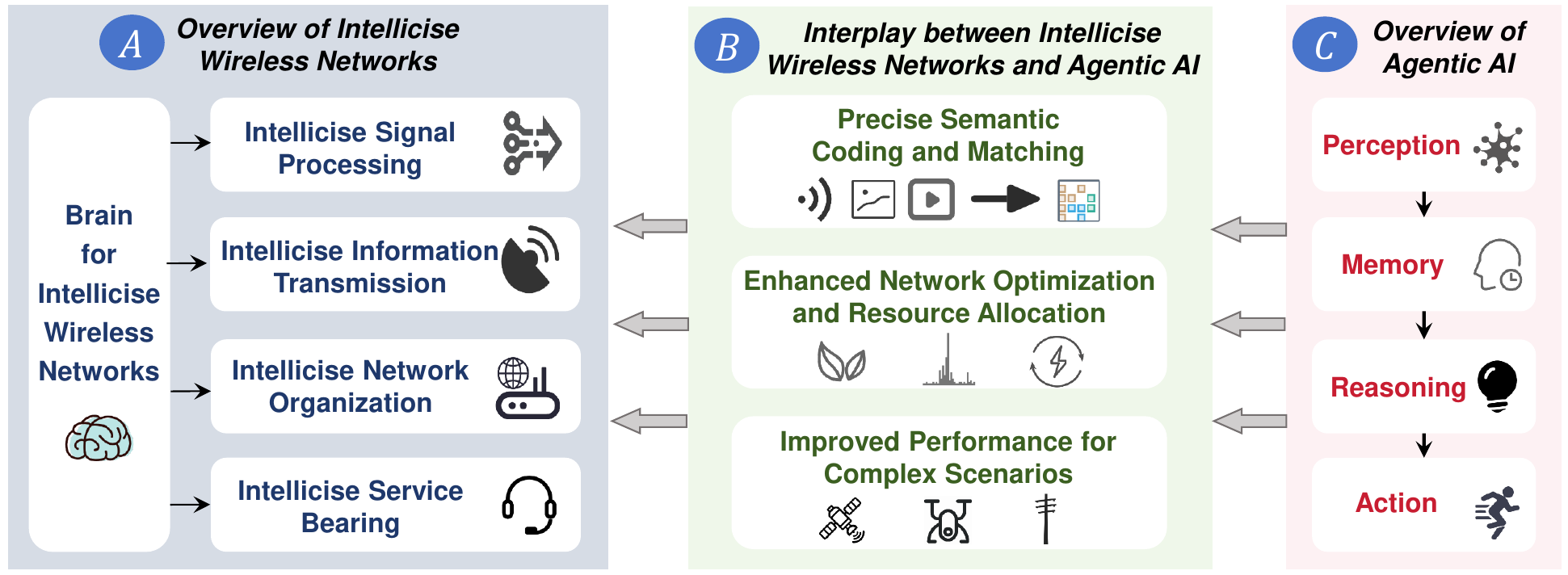}
\caption{Overview of Agentic AI-enhanced Intellicise Wireless Networks, where \textit{Part (A)} denotes the intellicise wirreless network, \textit{Part (C)} denotes the workflow of Agentic AI, and \textit{Part (B)} denotes three aspects of interplay between intellicise wireless networks and Agentic AI.}
\label{figure1}
\end{figure*}

\subsection{Overview of Intellicise Wireless Networks}


As illustrated in Fig. \ref{figure1} \textit{Part (A)}, the intellicise wireless network includes the following modules \cite{zhang2025intellicise}.

\begin{itemize}
\item \textbf{Brain for Intellicise Wireless Networks:} The brain serves as the central intelligence hub for intellicise wireless networks. By continuously ingesting and analyzing real-time data, the brain performs intelligent state inference to assess critical network parameters. Then, the system generates evolutionary strategies for network optimization. Furthermore, it translates these strategies into granular control directives across multiple network layers.
\item \textbf{Intellicise Signal Processing:} Guided by the brain, intellicise signal processing employs cutting-edge paradigms to revolutionize chip architectures, signal modeling, and processing algorithms. The integration of advanced signal processing techniques enables real-time resource optimization while maintaining operational simplicity, thereby supporting long-term network adaptability without compromising efficiency.
\item \textbf{Intellicise Information Transmission:} SemCom is positioned as a transformative technology for intelligent wireless networks, enhancing efficiency in both information processing and transmission. Guided by the brain, intellicise semantic information transmission dynamically adapts information representation granularity and encoding strategies to align with diverse transmission objectives and service requirements \cite{meng2025semantic}.
\item \textbf{Intellicise Network Organization:} Intellicise network organization comprises three interdependent pillars: super-wise nodes, extremely-flexible links, and ultra-concise networks, all orchestrated by the brain. Super-wise nodes act as autonomous decision-making units, extremely-flexible links enable dynamic network reconfiguration, and ultra-concise networks focus on structural and protocol-level optimization.
\item \textbf{Intellicise Service Bearing:} Intellicise service bearing operates as an adaptive service delivery engine guided by the brain. It investigates user intentions and service characteristics through dynamic analysis to support diverse service types. This component provides feedback to the brain, supplying essential data for intelligent network adjustments. Simultaneously, it adapts its service delivery mechanisms based on brain's control directives, enabling real-time optimization of service-bearing strategies.
\end{itemize}

\subsection{Overview of Agentic AI}

As illustrated in Fig. \ref{figure1} \textit{Part (C)}, Agentic AI includes the following workflow \cite{zhang2026toward}.
\begin{itemize}
\item \textbf{Perception:} The perception module is responsible for capturing and parsing multimodal information from the environment, providing diverse inputs for subsequent decision-making and action execution. Unlike traditional AI's passive response mechanism, Agentic AI's perception module actively integrates heterogeneous data to achieve environmental comprehension and intention inference, thereby supporting autonomous closed-loop task execution.
\item \textbf{Memory:} The memory module serves as a core component that transcends LLMs' context window limitations, enabling long-term autonomous operation and personalized interaction. It emulates human cognitive mechanisms through functionalities, including storage, retrieval, reflection, and reinforcement, thereby supporting dynamic information processing.
\item \textbf{Reasoning:} The reasoning module executes logical reasoning, causal analysis, decision formulation, and planning generation, driving the agent's closed-loop transition from the memory to action module. Its fundamental breakthrough lies in surpassing LLMs' passive generation capabilities, achieving goal-oriented active reasoning, multi-step planning, and decision-making under uncertainty, thus supporting autonomous completion of complex tasks.
\item \textbf{Action:} The action module translates decisions generated by the reasoning module into concrete actions, enabling interaction with external environments to accomplish goal-oriented tasks. These actions encompass multi-agent collaboration, robotic and device control, UI and system automation, tool invocation, and other operational implementations.
\end{itemize}

\subsection{Interplay between Intellicise Wireless Networks and Agentic AI}

As illustrated in Fig. \ref{figure1} \textit{Part (B)}, the principal conjunctions between intellicise wireless networks and Agentic AI with the corresponding elaboration are listed as follows.
\begin{itemize}
\item \textbf{Precise Semantic Coding and Matching:} The perception module captures multimodal data including text, images, and voice, and integrates it with the semantic knowledge base stored in the memory module. Through logical reasoning and intent inference, the reasoning module achieves precise semantic matching. Additionally, the memory module continuously updates the semantic knowledge base using RAG techniques, while the reasoning module optimizes decision-making strategies via DRL, enabling adaptive semantic coding in dynamic environments \cite{jiang2025large}.
\item \textbf{Enhanced Network Optimization and Resource Allocation:} The perception module monitors real-time network status, and the reasoning module performs resource requirement and energy efficiency analysis. The action module then executes optimization strategies, driving self-optimization of spectral efficiency, energy efficiency, and operational efficiency. This closed-loop process ensures adaptive resource management and optimization, aligning with the demands of intellicise wireless networks \cite{liu2025lameta}. 
\item \textbf{Improved Performance for Complex Scenarios:} 
In complex scenarios like autonomous driving systems and space-air-ground integrated networks, Agentic AI shows exceptional capability in perceiving multi-dimensional, cross-domain, and cross-modal environmental data. It then systematically organizes, compresses, and retrieves stored knowledge with high efficiency to support reasoning. Furthermore, it leverages multi-agent collaboration technologies to enable precise task decomposition, adaptive strategy optimization, and the execution of optimal actions, thereby significantly improving system robustness and decision-making quality in dynamic scenarios \cite{zhang2025embodied}.
\end{itemize}



\begin{figure*}
\centering
\includegraphics[width=1\textwidth]{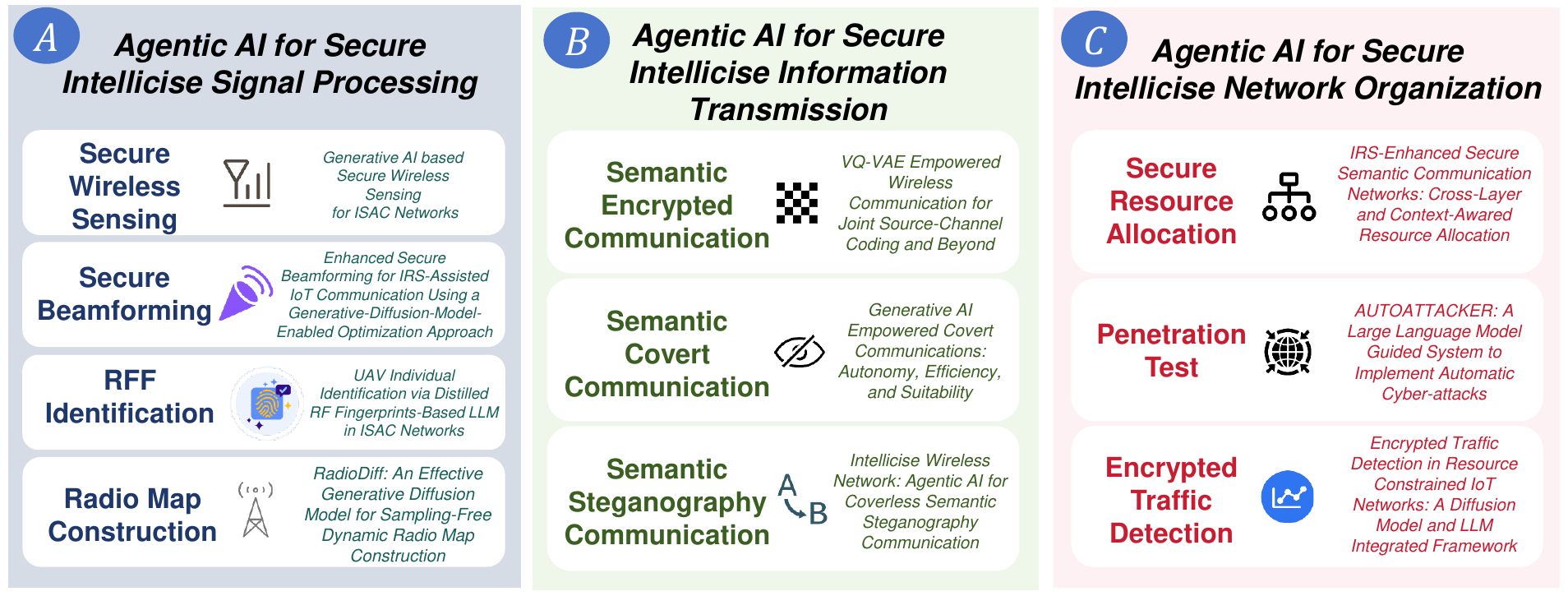}
\caption{Illustration of Agentic AI-enhanced secure intellicise wireless networks, where the italic text is the title of the corresponding article. \textit{Part (A)} denotes Agentic AI for secure intellicise signal processing, \textit{Part (B)} denotes Agentic AI for secure intellicise information transmission, and \textit{Part (C)} denotes Agentic AI for secure intellicise network organization.}
\label{figure2}
\end{figure*}

\section{Agentic AI for Secure Intellicise Wireless Networks}
\label{section3}

In this section, we propose a detailed taxonomy for Agentic AI-enhanced secure intellicise wireless networks, as shown in Fig. \ref{figure2}. 

\subsection{Agentic AI for Secure Intellicise Signal Processing}

\subsubsection{Agentic AI for Secure Wireless Sensing}

As a prime illustration of Integrated Sensing and Communication (ISAC), CSI-based wireless sensing enhances spectral efficiency, energy efficiency, and operational cost savings while minimizing hardware expenditures for intellicise wireless networks. Yet, any conventional wireless device can analyze CSI samples captured in open environments to infer private details about individuals' daily activities. To address this security vulnerability, Agentic AI dynamically generates a protective signal and modulates it onto the training symbols employed for CSI estimation. This intervention effectively masks signal fluctuations induced by user activities, preserving privacy without compromising sensing performance.

\subsubsection{Agentic AI for Secure Beamforming}
Beamforming achieves directional signal transmission by precisely adjusting the phase and amplitude of antenna arrays, thereby enhancing the coverage capability of intellicise wireless networks. However, the spatial correlation between legitimate and unauthorized channels introduces significant security challenges for beamforming. To overcome this, Agentic AI derives optimal secure beamforming strategies from noisy multi-user wireless channel environments, effectively improving the secrecy rate in dynamic scenarios.

\subsubsection{Agentic AI for Radio Frequency Fingerprint identification}

Radio Frequency Fingerprint (RFF) identification delivers a lightweight, safe-endogenous identity recognition mechanism for intellicise wireless networks. By integrating Agentic AI, this approach combines pre-trained LLMs with dynamic knowledge distillation techniques to refine identification accuracy in complex environments. Additionally, it efficiently transfers learned knowledge to lightweight models, optimizing deployment efficiency while maintaining robust performance.

\subsubsection{Agentic AI for Radio Map Construction}

Electromagnetic signal sensing and radio map reconstruction empower the intellicise wireless network to pinpoint malicious or active eavesdroppers tied to suspicious signal sources within the radio map. Integrating Agentic AI, the system can further leverage base station positions and environmental characteristics as contextual prompts, enabling efficient construction of sampling-free radio maps while enhancing detection precision in dynamic environments.

\subsection{Agentic AI for Secure Intellicise Information Transmission}

\subsubsection{Agentic AI for Semantic Encrypted Communication}

By leveraging codebook-driven noise modeling, distortion-aware training, and dynamic dependency adaptation, Agentic AI fortifies semantic encrypted communications. Unauthorized parties lacking access to the specific codebook cannot decipher private semantics, as the system intentionally obscures semantic mappings. Additionally, Agentic AI jointly optimizes the semantic encoder, decoder, and codebook, tailoring each component to dataset characteristics and scenario-specific requirements.

\subsubsection{Agentic AI for Semantic Covert Communication}

Semantic covert communication defends against eavesdroppers by minimizing detection and interception risks. Traditional approaches like frequency hopping and spread spectrum, however, are becoming increasingly ineffective as adversaries enhance their detection and decryption capabilities. Agentic AI, leveraging self-supervised and continuous learning for adaptability, demonstrates exceptional versatility in dynamic environments, rendering it uniquely adept at optimizing covert strategies. By capitalizing on this adaptability, it dynamically adjusts key parameters, including power allocation, frequency selection, and modulation schemes, to optimize semantic covert transmission across evolving threat landscapes.

\subsubsection{Agentic AI for Semantic Steganography Communication}

Steganography embeds confidential information within cover modality data, making it so that eavesdroppers can only reconstruct the cover's semantic features. Distinct from cryptographic techniques that explicitly encrypt semantic data, steganography achieves three critical advantages: it conceals the encryption process itself, resists detection by intelligent eavesdroppers, and synergistically integrates with traditional encryption to establish dual-layer protection.
Notably, Agentic AI enables ceverless semantic steganography communication through adaptive and conditional generation. This breakthrough addresses three fundamental limitations: first, it circumvents the capacity constraints imposed by cover images; second, it reduces the risk of anomaly detection through cover image analysis; and third, it enhances resilience against advanced steganalysis techniques \cite{meng2025image}.

\subsection{Agentic AI for Secure Intellicise Network Organization}

\subsubsection{Agentic AI for Secure Resource Allocation}

Agentic AI adopts a dual-layer architecture integrating short-term and long-term memory to construct a dynamically adjustable, semantically context-aware state space. This space comprises both a rich semantic subspace and an observable environmental state subspace. Through joint optimization of semantic representation resources, sub-channel allocation, computational capacity, and perceptual modalities, Agentic AI effectively captures semantic contexts while addressing high-dimensional space learning challenges. This synergistic optimization ultimately elevates both secure semantic transmission rates and secure semantic spectral efficiency.

\subsubsection{Agentic AI for Penetration Test}


Penetration test actively uncovers and exploits network vulnerabilities to verify its security protection capabilities, ultimately outputting remediation recommendations to enhance network security. Agentic AI can dynamically plan attack paths, autonomously adjust attack strategies, and reduce human intervention, thus improving testing efficiency. Additionally, by leveraging RAG technology, Agentic AI can enhance its knowledge base with experiences from previous attack actions before generating the next attack move. This increases the likelihood of successful attacks and introduces more network vulnerability samples for securing intellicise wireless networks.

\subsubsection{Agentic AI for Encrypted Traffic Detection}


Encrypted traffic detection technology identifies, analyzes, and monitors data traffic transmitted through encrypted protocols to uncover potential security threats or abnormal behaviors. Introducing Agentic AI brings the following advantages:
Firstly, Agentic AI can overcome noise characteristics in network data to extract discriminative features from visual representations of network traffic, particularly in encrypted traffic scenarios with ambiguous patterns.
Secondly, Agentic AI enables efficient exploration of complex and high-dimensional search spaces for capturing fine-grained and abstract traffic.
Finally, Agentic AI maintains robust detection performance under resource constraints while adapting to evolving data distributions.

\section{Securing Agentic AI-enhanced Intellicise Wireless Networks}

\begin{table*}
  \centering 
  \caption{The security and privacy threats during the four stages of Agentic AI.}
  \label{attack}
  \renewcommand{\arraystretch}{1.5}
  
  \newcommand{\gc}{\cellcolor{gray!15}}
  \newcommand{\tabhead}[1]{\multicolumn{1}{|c|}{\textbf{#1}}}

  \begin{tabular}{|>{\centering\arraybackslash}m{1.5cm}|
                >{\centering\arraybackslash}m{3cm}|
                >{\raggedright\arraybackslash}m{11.5cm}|}
    \hline
    \tabhead{Stage} & \tabhead{Security Threat} & \tabhead{Description} \\ 
    \hline
    
    \multirow{5}{*}{Perception} 
      & \gc Injection Attack & \gc Inserting multi-modal instructions or masking intent to bypass defenses, while utilizing fragmented injection or single-node infection to trigger systemic chain reactions. \\
    \cline{2-3} 
      & Perception Vulnerability & Terminating processing prematurely, mapping intents inaccurately in ambiguous scenarios, and interfering with internal templates to disable environmental recognition. \\
    \cline{2-3}
      & \gc Inference Attack & \gc Reverse inference internal states and prompts by observing behavioral patterns and feedback during continuous interactions. \\
    \cline{2-3}
      & Localized Mismatches & Inducing localization cognitive biases by failing to adapt to regional differences such as time zones and languages. \\
    \cline{2-3}
      & \gc Robot Detection & \gc Imposing reverse constraints by misidentifying legitimate agents as malicious bots to block autonomous perception and deny service access. \\
    \hline 

    Memory 
      & Poisoning Attack & Poisoning the shared knowledge base and historical memory to inject persistent cognitive biases, manipulating autonomous evolution and causing systematic network failures. \\
    \hline

    \multirow{4}{*}{Reasoning} 
      & \gc Path Hijacking & \gc Hijacking the reasoning path to manipulate the logic chain or launching jailbreak attacks to bypass security reviews. \\
    \cline{2-3}
      & Objective Function Corruption & Corrupting objective functions and reward weights, tampering with confidential decision data, and using maliciously reinforced beliefs to distort network topology and semantic evolution. \\
    \cline{2-3}
      & \gc Multi-Agent Collusion & \gc Creating an erroneous group consensus through multi-agent collusion and interrupting logic chains prematurely to sacrifice reasoning depth, leading to the misjudgment of unfeasible network states. \\
    \cline{2-3}
      & Protocol-Level Attack & Exploiting model-specific protocols to enable deceptive interaction logic and induce unintended malicious behaviors during instruction processing. \\
    \hline

    \multirow{4}{*}{Action} 
      & \gc Autonomy Abuse & \gc Manipulating tool-invocation permissions to break operational boundaries, exploiting perceptual blind spots to induce destructive adjustments, and launching denial-of-service attacks by designing high-computational-cost invalid actions. \\
    \cline{2-3}
      & Oversight Saturation & Generating a flood of low-importance audit events to exhaust monitoring resources. \\
    \cline{2-3}
      & \gc Governance Obfuscation & \gc Exploiting complex interaction mechanisms and log ambiguity to obscure responsibility trails, making actions unattributable and enabling long-term lurking. \\
    \hline

  \end{tabular}
\end{table*}

\subsection{Security and Privacy Issues Brought by Agentic AI}

Table \ref{attack} summarizes the security and privacy threats that occur in the four stages of Agentic AI, and their impact on intellicise wireless networks is described below.

\subsubsection{Security and Privacy Threats in the Perception Stage}
The threats compromise the semantic perception and decision-making integrity of Agentic AI, leading to the execution of malicious commands and the failure of the network to accurately understand or adapt to its environment.
Firstly, injection attacks introduce malicious command execution, defensive bypasses, and cross-node infection, potentially leading to widespread system compromise or operational failure within the intellicise wireless network through both intentional multi-modal injections and unintentional user errors.
Secondly, perception vulnerabilities compromise the reliability of the intellicise wireless network's decision-making core by causing incomplete information processing, inaccurate intent mapping, and the failure of environment-sensing templates under adversarial interference.
Thirdly, during agents’ continuous exploration and perception, inference attacks can reverse infer agents’ internal states and prompt information by observing their perceptual behaviors and feedback patterns.
Fourthly, localization mismatches in time and language create contextual cognitive biases, undermining the intellicise wireless network's ability to maintain accurate situational awareness and consistent service performance across diverse geographic regions.
Fifthly, robot detection may cause service denial and functional paralysis by misidentifying legitimate autonomous agents as malicious bots, severing their ability to perceive the environment or access essential intellcise wireless network services \cite{datta2025agentic}.

\subsubsection{Security and Privacy Threats in the Memory Stage}
Attackers inject misleading information, which slowly infiltrates the knowledge base, contextual memory, and policy update mechanism of the intellicise Wireless Network. This poisoning attack can cause the network to continuously incorporate tampered data into its autonomous evolution process under seemingly normal operating conditions, resulting in long-term erroneous decisions. In addition, it can even use legitimate permissions to execute malicious instructions \cite{narajala2025securing}.

\subsubsection{Security and Privacy Threats in the Reasoning Stage}
The threats compromise the logic-based global control and collaborative reasoning integrity of the intellicise wireless network, leading to hijacked decision chains, corrupted evolution goals, and the collapse of trustworthy group consensus.
Firstly, path hijacking compromises the logic-based global control and secure decision-making integrity of the intellicise wireless network, enabling attackers to hijack reasoning paths and bypass operational constraints to execute unauthorized network configurations.
Secondly, objective function corruption and multi-agent collusion compromise the self-evolution and objective-driven reasoning of the intellicise wireless network, leading to a permanent deviation of the system’s long-term goals, network topologies, and semantic integrity.
Thirdly, protocol-level attacks compromise the inter-node collaborative reasoning and logical foundations of the intellicise wireless network by exploiting communication protocols to inject deceptive interaction logic and trigger unintended malicious behaviors \cite{datta2025agentic}.




\subsubsection{Security and Privacy Threats in the Action Stage}
The threats compromise operational boundary control, security oversight, and the integrity of the trust ecosystem, leading to unauthorized actions, governance evasion, and the collapse of identity-based collaborative security.
Firstly, autonomy abuse compromises operational authority control and resource management efficiency, leading to the intellicise wireless network's unauthorized system reconfigurations and the exhaustion of computing resources through malicious tool invocation.
Secondly, oversight saturation compromises the security monitoring and incident response efficiency of the intellicise wireless network by using a large number of low-priority audit events, alerts, or operations requiring manual review to obscure critical attack detection.
Thirdly, governance obfuscation allows attackers to obscure their responsibility trails and maintain undetected, long-term persistence within the intellicise wireless network \cite{raza2025trism}.

\subsection{Security and Privacy Protection Techniques}

Table \ref{defense} summarizes the defense technologies for addressing security and privacy threats introduced by Agentic AI, as described in detail below.

\begin{table*}
  \centering 
  \caption{The defense techniques for addressing security and privacy threats introduced by Agentic AI.}
  \label{defense}
  \renewcommand{\arraystretch}{1.5}
  
  \newcommand{\gc}{\cellcolor{gray!15}}
  \newcommand{\tabhead}[1]{\multicolumn{1}{|c|}{\textbf{#1}}}

  \begin{tabular}{|>{\centering\arraybackslash}m{2cm}|
                >{\centering\arraybackslash}m{3cm}|
                >{\raggedright\arraybackslash}m{11cm}|}
    \hline
    \tabhead{Category} & \tabhead{Protection Technique} & \tabhead{Description} \\
    \hline
    
    \multirow{8}{2cm}{\centering Robustness Enhancement} 
      & \gc Adversarial Training & \gc Implement adversarial training by proactively injecting disturbances into encoders and decoders to minimize sensitivity to interference. \\
    \cline{2-3} 
      & Red-Team Testing & Utilize red team testing to simulate extreme pressure and verify maximum system capacity. \\
    \cline{2-3}
      & \gc Instruction Hierarchy & \gc Implement a strict instruction hierarchy to prioritize security constraints over operational intents. \\
    \cline{2-3}
      & Supervised Fine-Tuning & Apply supervised fine-tuning using curated datasets to recalibrate model parameters and embed hard constraints into decision-making logic. \\
    \cline{2-3}
      & \gc Integrity Verification & \gc Perform rigorous validation of data sources and parameters to ensure use of untampered information in decision-making. \\
    \cline{2-3}
      & Cryptography Technology &  Protect data or parameters from eavesdropping through cryptographic encryption, differential privacy and other means.\\
    \cline{2-3}
      & \gc Cross Checking & \gc Deploy dedicated detection models and implement inter-agent cross-check mechanisms to identify abnormal traffic and behaviors in real-time. \\
    \cline{2-3}
      & Data Minimization & Apply data minimization to filter redundant information, while integrating real-time quality preprocessing and feedback loops to remove impurities from the knowledge flow. \\
    \hline 
    
    \multirow{5}{2cm}{\centering Trustworthy Design} 
      & \gc Universal Security Architecture & \gc Internalize industry security standards as mandatory guidelines to establish a standardized risk assessment system and govern interaction logic across all nodes and devices. \\
    \cline{2-3}
      & Protocol-Based Communication & Enforce protocol-based semcom to standardize the communication interface and block non-standard protocols, effectively preventing out-of-band attacks and logical obfuscation. \\
    \cline{2-3}
      & \gc Trusted Execution Environments & \gc Utilize trusted execution environments to isolate critical reasoning and load code/data into hardware enclaves, ensuring physical separation from the host operating system. \\
    \cline{2-3}
      & Double Verification  & Enforce a dual verification mechanism and utilize multi-party consensus protocols to obtain secondary authorization for high-risk operations and validate critical instructions. \\
    \cline{2-3}
      & \gc Context-Aware Authentication & \gc Analyzing multi-dimensional parameters in real-time to dynamically verify collaboration requests alongside traditional credentials. \\
    \hline
    
    \multirow{5}{2cm}{\centering Real-Time Defense} 
      & Anomaly Detection & Deploy anomaly detection to perform continuous monitoring of data and parameters and integrate sensitive data identification to trigger immediate alerts upon detecting deviations. \\
    \cline{2-3}
      & \gc Input Optimization & \gc Standardize interfaces through structured queries, apply prompt cleaning and ignoring strategies, enforce strict filtering to block injection attacks. \\
    \cline{2-3}
      & Multi-Layered Isolation & Construct a multi-layered sandbox architecture to enforce distributed runtime execution and confine operational access for unverified nodes within isolated environments. \\
    \cline{2-3}
      & \gc Rollback Mechanism & \gc Configure an automated loss-stop mechanism by integrating a runtime termination switch and executing a state restoration sequence to rollback configurations to a stable baseline. \\
     \cline{2-3}
      & Audit Log System & Establish a detailed audit log system and integrate blockchain technology to enforce log immutability and create permanent records for forensic review and fault localization. \\
    \hline
  \end{tabular}
\end{table*}

\subsubsection{Robustness Enhancement}
Robustness enhancement techniques elevate the overall robustness of the intellicise wireless network through targeted training optimization and defensive measures, securing the system’s anti-interference capability, operational stability, and reliable operation in complex scenarios \cite{datta2025agentic,narajala2025securing,raza2025trism,shahriar2025survey,beurer2025design}.
 
\begin{itemize}
\item \textbf{Adversarial Training:} It enhances the resilience against malicious features, ensuring the system can maintain high-fidelity semantic processing and stable reasoning logic even when subjected to malicious interference. 
\item \textbf{Red Team Testing:} It can simulate high-intensity external pressure to proactively verify the system’s maximum bearing capacity, ensuring that intellicise wireless networks possess self-diagnostic and rapid response capabilities when confronting unknown threats.
\item \textbf{Instruction Hierarchy:} It protects the intellicise wireless network by effectively neutralizing hazardous commands, ensuring that core security principles remain inviolable even when the network is subjected to contradictory or high-risk inputs.
\item \textbf{Supervised Fine-Tuning:} It safeguards the intellicise wireless network by aligning autonomous decision-making with real-time optimization, which prevents unintended system divergence.
\item \textbf{Integrity Verification:} By maintaining strict oversight over knowledge updates and parameter sharing, it guarantees that all decisions are derived from unaltered information sources, effectively preventing threats caused by data tampering or corruption.
\item \textbf{Cryptography Technology:} It blocks information leakage risks at the source by encrypting signals and data transmitted over intellicise wireless networks, making it impossible for eavesdroppers to extract sensitive data.
\item \textbf{Cross Checking:} It protects the network from deceptive behaviors, and ensures that malicious activities are promptly identified, thereby maintaining the robustness in large-scale environments.
\item \textbf{Data Minimization:} By filtering out redundant data and continuously purifying knowledge flows, it protects the intellicise wireless network from the risks associated with processing unnecessary information.
\end{itemize}

\subsubsection{Trustworthy Design}
Trustworthy techniques enable reliable trust verification and secure collaborative interaction to fend off unauthorized access and collaboration-related risks \cite{datta2025agentic,narajala2025securing,raza2025trism,shahriar2025survey}.
\begin{itemize}
\item \textbf{Universal Security Architecture:} It ensures secure interoperability between diverse network components, and effectively mitigates systemic collaboration risks during complex, cross-domain interactions.
\item \textbf{Protocol-based Communication:} By eliminating ambiguity and enforcing strict logical rules, it prevents out-of-band attacks and semantic obfuscation, providing a reliable foundation for coordinated actions among distributed intellicise nodes.
\item \textbf{Trusted Execution Environment:} It ensures that critical reasoning is shielded from external interference, thereby maintaining physical execution integrity.
\item \textbf{Double Verification:} By establishing multi-party consensus requirements and conducting double verification for high-risk operations, the probability of single point failures can be greatly reduced.
\item \textbf{Context-aware Authentication:} It protects the intellicise wireless network against sophisticated spoofing and unauthorized access, ensuring that only legitimate requests are accepted within highly mobile and changing network environments.
\end{itemize}

\subsubsection{Real-Time Defense}
Real-time defense techniques enable timely threat detection, rapid risk response, and full-process traceability to ensure the network’s secure and stable operation \cite{datta2025agentic,narajala2025securing,raza2025trism,shahriar2025survey}.
\begin{itemize}
\item \textbf{Anomaly Detection:} By monitoring semantic inputs and model response logic, it can immediately alert and block potential semantic contamination or logical failures upon detecting any abnormal operation from expected patterns.
\item \textbf{Input Optimization:} It enforces strict compliance on all decision paths and neutralizing prompt-based threats, ensuring that interactions remain within safe boundaries.
\item \textbf{Multi-layered Isolation:} It confines threats to local scopes to avoid malicious agents from spread within intellicise wireless network, preventing single-point compromises from triggering a total network collapse.
\item \textbf{Rollback Mechanism:} It ensures core service continuity by providing an automated safety network that recovers the intellicise wireless network from severe security impacts.
\item \textbf{Audit Log System:} It safeguards the intellicise wireless network by enabling rapid fault localization and forensic analysis, ensuring every decision can be audited to maintain a verifiable and secure operational environment.
\end{itemize}

\section{Case Study: Agentic AI for Semantic Steganography Communication}

\begin{figure*}
\centering
\includegraphics[width=1\textwidth]{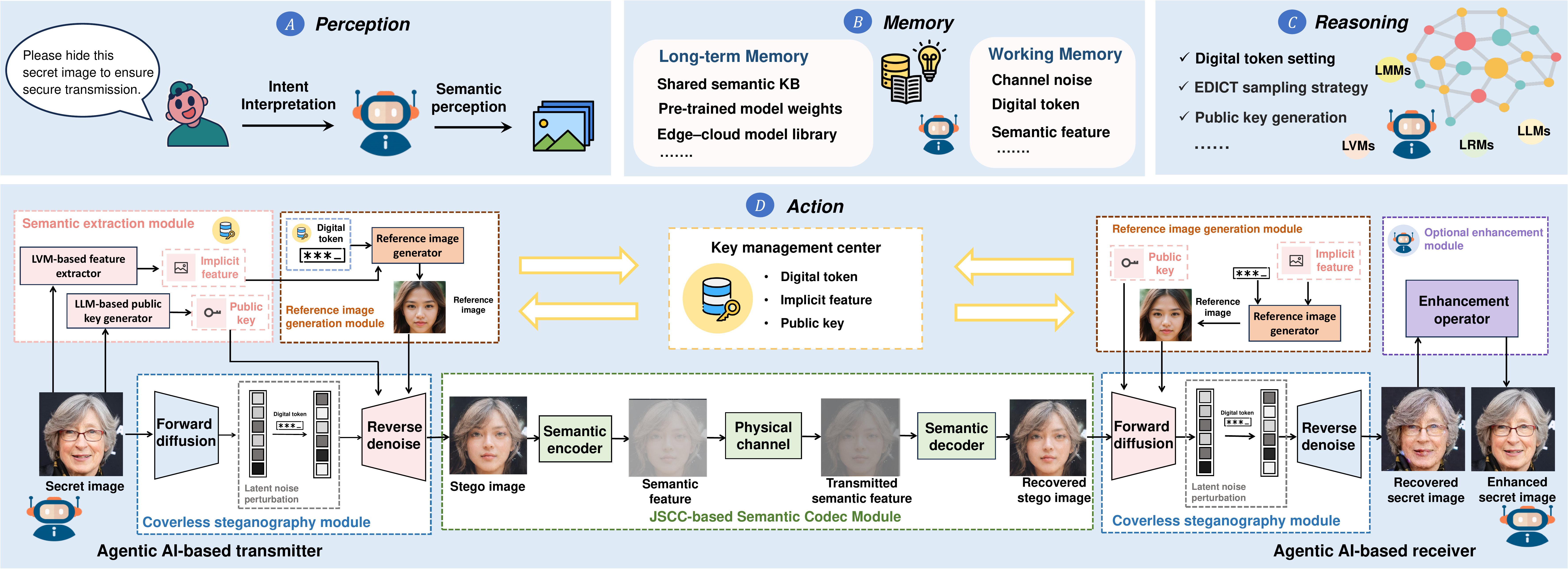}
\caption{Illustration of the presented Agentic AI-based semantic steganography communication scheme, where \textit{Part (A)} represents the perception stage for perceiving secure transmission requirements and multi-modal semantic source information; \textit{Part (B)} represents the memory stage for semantic model sharing and updating; \textit{Part (C)} represents the reasoning stage for key generation and digital token setting; and \textit{Part (D)} represents the action stage for the implementation of the semantic steganography strategy and optional semantic enhancement.}
\label{figure3}
\end{figure*}

In this section, we present an application of Agentic AI specifically tailored for semantic steganography communication in intellicise wireless networks.

\subsection{Challenges}

Although generative AI-based semantic steganography communication achieves superior steganographic capacity and eliminates reliance on cover images compared to cover image-based methods \cite{meng2025image}, it still encounters the following challenges:
\begin{itemize}
\item \textbf{Private Semantic Information Leakage:} The semantic text private key and private image maintain semantic consistency, rendering them vulnerable to cracking under specific domain knowledge contexts. This creates a pathway for adversarial extraction of confidential semantic content.
\item \textbf{Frequent Private Key Updates:} Each private image requires a dedicated private key, necessitating key updates with every transmission. This not only increases complexity but also heightens the risk of private information leakage.
\end{itemize}

\subsection{Presented Scheme}

As illustrated in Fig. \ref{figure3}, the presented Agentic AI-based semantic steganography communication scheme includes the following steps.
\begin{itemize}
\item \textbf{Step 1: Multi-modal Semantic Perception.} By utilizing Large Vision Models (LVMs), Agentic AI extracts semantic segmentation or skeleton graph, perceiving implicit features required for semantic steganography.

\item \textbf{Step 2: Key and Token Configuration.} Agentic AI generates public semantic keys through LLMs. Notably, public keys carry distinct semantic meanings from the corresponding secret image.
To control reference image generation, Agentic AI employs digital tokens to determine the initial noise and injects implicit feature through ControlNet \cite{yang2024diffstega}.

\item \textbf{Step 3: Stego Image Generation and Transmission.} 
Leveraging EDICT, Agentic AI generates stego image under the conditions of the public key and reference image. During this process, deterministic perturbations governed by the digital tokens are applied to a noisy latent vector.
Agentic AI further chooses appropriate semantic codec for semantic transmission of the stego image.

\item \textbf{Step 4: Secret Image Recovery and Enhancement.} 
At the receiver, Agentic AI decodes the stego image based on the shared semantic knowledge base, and recovers the secret image using the digital token. Agentic AI can further invoke specialized tools from the memory module for task-oriented enhancement, such as super-resolution model for facial recognition.
\end{itemize}

\subsection{Results and Analysis}
\begin{figure}
\centering
\includegraphics[width=0.45\textwidth]{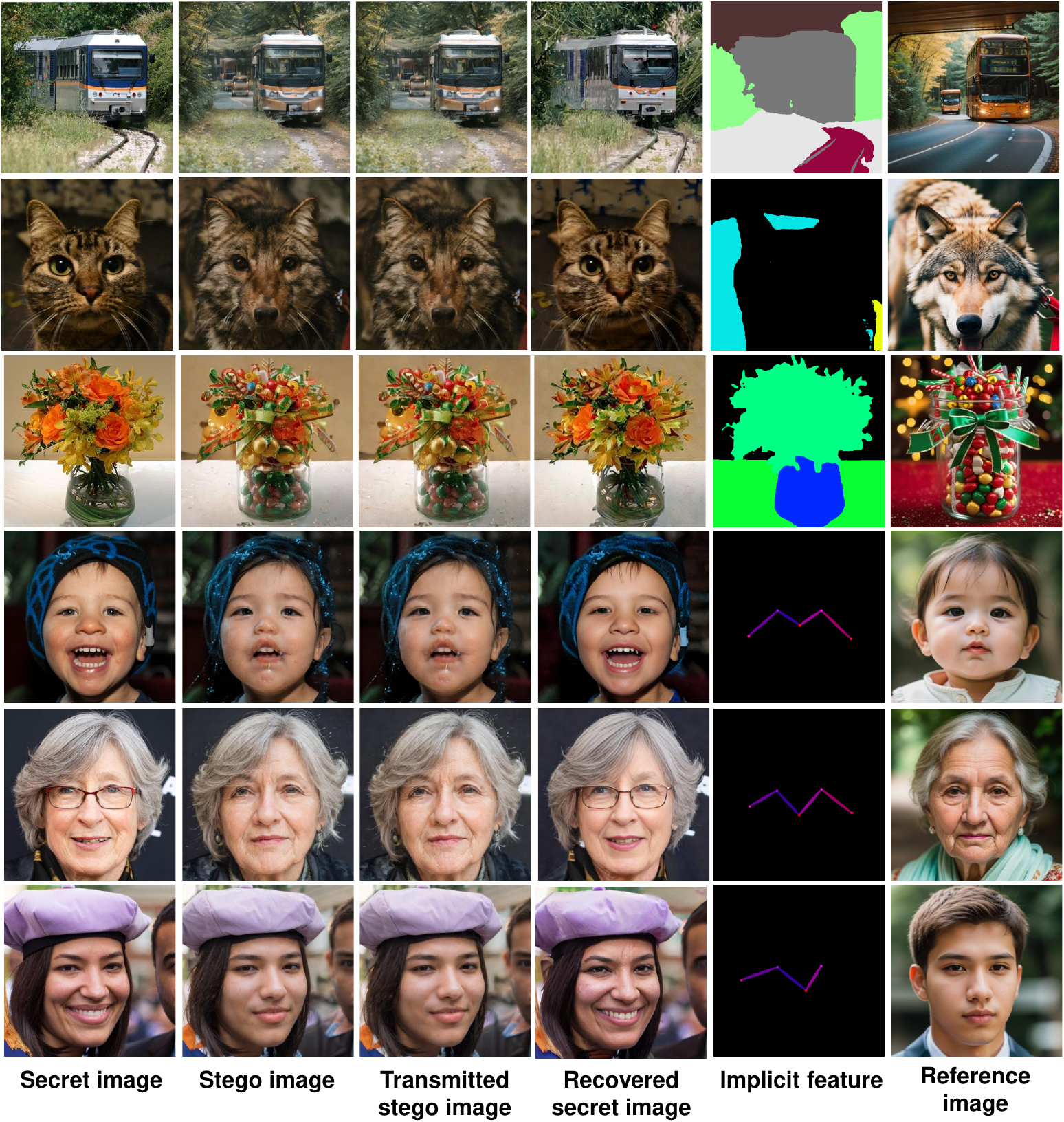}
\caption{Simulation results of the case study. We select the Stable Diffusion version 1.5 as the conditional diffusion model and employ the EDICT as the sampling algorithm. Both the forward and reverse processes are configured to include 50 steps. The SwinJSCC architecture is employed as the trained semantic encoder and decoder.}
\label{figure4}
\end{figure}

Fig. \ref{figure4} provides visualization results of the presented scheme under the UniStega dataset \cite{yang2024diffstega}. The legitimate receiver applying the correct digital token can generate correct reference image to further recover the secret image, while the eavesdropper only intercepts stego image with wrong content. Even eavesdroppers try to recover the secret information with the public semantic key and implicit feature, the deterministic perturbations and exactly inverse sampling controlled by digital token prevent the correct reconstruction, which causes the image collapsion or misleading contents.


\section{Open Research Issues and Outlooks}

In this section, we outline key issues and future perspectives to guide research directions in Agentic AI-enhanced secure intellicise wireless networks.

\subsection{Agentic AI for Secure and Trustworthy Network Architecture}
As described in Section \ref{section3}, existing solutions focus on utilizing Agentic AI to enhance modular security capabilities, such as secure wireless sensing and secure beamforming. However, designing a secure and trustworthy network architecture based on Agentic AI to achieve higher-level defense capabilities remains a challenge. Future research can focus on Agentic AI-enhanced endogenous-security network architecture, firewalls, and intrusion detection and defense systems \cite{liu2025secure}.

\subsection{Low Complexity Agentic AI Model Design}
The deployment of Agentic AI in intellicise wireless networks requires consideration of security, service performance, and energy consumption to achieve optimal trade-offs. It is crucial to design low complexity Agentic AI models while ensuring the security of networks on end users with insufficient computing power. Future research directions can focus on compression techniques for Agentic AI models, including low-rank adaptation, knowledge distillation, quantification, and pruning \cite{zhang2026toward}.


\subsection{Platform Establishment and Experimental Test for Secure Intellicise Wireless Networks}
Establishing a comprehensive experimental platform is critical for validating theoretical and technological advancements of Agentic AI-enhanced intellicise wireless networks. Future efforts should prioritize: 1) Defining standardized metrics for security and privacy protection, enabling systematic comparison of different Agentic AI implementations; 2) Developing a simulation environment integrated with physical testbeds to validate security protocols under both controlled laboratory conditions and realistic wireless channel impairments; and 3) Conducting field trials in typical scenarios to assess the security capability against real-world threats.

\section{Conclusions}
In this paper, we have first reviewed intellicise wireless networks and Agentic AI, followed by the interplay between them. Then, we have proposed a taxonomy for Agentic AI-based secure intellicise wireless networks, including secure intellicise signal processing, secure intellicise information transmission, and secure intellicise network organization. Moreover, we have summarized the security and privacy issues introduced by Agentic AI in intellicise wireless networks, and further identified corresponding defense technologies. Additionally, we have given a case study to show the potential of Agentic AI in protecting intellicise wireless networks from attacks by intelligent eavesdroppers. Lastly, we have provided open research issues and outlooks.

\bibliography{ref.bib}
\bibliographystyle{IEEEtran}

\vfill

\end{document}